\documentclass{aa} 
\usepackage{graphicx}
\usepackage{txfonts}
\usepackage{color}
\usepackage{float}
\newcommand{\acena}{$\alpha \, {\rm Cen\,A}$}          
\newcommand{\acenb}{$\alpha \, {\rm Cen\,B}$}                          %
\newcommand{\acen}{$\alpha \, {\rm Cen}$}                              %
\newcommand{\about}{$\sim$}                       
\newcommand{\bpic}{$\beta \, {\rm Pic}$}

\newcommand{\amin}{$^{\prime}$}                   
\newcommand{\asec}{$^{\prime \prime}$}
\newcommand{\adeg}{$^{\circ}$}

\newcommand{\asecdot}[2]{\mbox{#1$\stackrel {\prime \prime}{_{\bf \cdot}}$#2}}

\begin{document} 

\title{{\bf $\alpha$}\,Centauri revisited: 2nd epoch ALMA observations}

   \author{
                                R. Liseau\inst{}                      
          }

   \institute{Department of Earth and Space Sciences, Chalmers University of Technology, Onsala Space Observatory, SE-439 92 Onsala, Sweden,
                                 \email{rene.liseau@chalmers.se}                                                                                                                                                    
           }
                
   \date{Received ; accepted}
  \abstract
   {The observational study of stars in the sub-millimetre regime has only rather recently begun and was made possible mainly by the Atacama Large Millimeter/submillimeter Array (ALMA). The emission mechanisms of this radiation from normal Main-Sequence stars and its physical significance for the outer atmospheric layers is the topic of intense contemporary study.}
   {Our previous ALMA observations of the \acen tauri binary system detected the submm emission originating in the chromospheres of these solar-type stars. Observations at another epoch are aiming at further characterising these atmospheric layers and their behaviour with time. In addition, we were aiming at clarifying the status of the recently discovered U source and its relation to the \acen\ system.}
   {The comparison of data from two epochs should present the basis for more advanced theoretical modelling of the chromospheres of \acen A and B. Proper motion data of the U source should establish its relation to the \acen\ system, and U's submm spectral energy distribution (SED) should provide information about its physical nature.}
   {In Cycle\,4, both stars were again detected in the same bands as in the earlier Cycle\,2. These early data suggested a flattening of the SED towards longer wavelengths. By analogy with the Sun, this was not expected. Eventually, it turned out to be caused by an obsolete calibration, but this has now been remedied. Each SED exhibits now a single spectral slope over the entire frequency range (90 to 675\,GHz). For the U source, the upper limits on its proper motion (pm) are much smaller than the pm of \acen, which essentially excludes any physical relationship with the binary.}
   {The second epoch ALMA observations of \acen\  did not confirm the flattening of the SED in the lowest frequency bands that was reported before. Rather, this was the result of an inadequate flux calibration using the minor planet Ceres. Over the entire frequency range observed with ALMA, the SEDs from Cycle\,4 can be fit by power laws of the form $S_{\!\nu} \propto \nu^{\,\alpha}$ with $\alpha = 1.76 \pm 0.01$ for \acen\,A and $\alpha = 1.71 \pm 0.02$ for \acen\,B. For the infrared/submm background object U applies $\alpha = 2.55 \pm 0.14$. If this emission from U is due to dust, its opacity exponent $\beta = \alpha - 2$ would be about 0.5, indicative of particle sizes that are larger than those of the interstellar medium ($\beta_{\rm \,ISM} \sim 2$), but comparable to those found in circumstellar discs.}
   \maketitle
%

\section{Introduction}

During ALMA-Cycle\,2, Alpha Centauri (\acen) has been observed in 2014 and 2015. The results of these successful observations have been communicated by \citet{liseau2015} and \citet{liseau2016}. For the further study of the chromospheric emissions from the binary stars, a new observing campaign was initiated during Cycle\,4 (2016 - 2017), again exploiting all of the previously available frequency bands, i.e. bands\,3, 4, 6, 7, 8 and 9 in the range 90 to 675\,GHz (3 to 0.4\,mm). This generated second epoch datasets for the study of the chromospheres of the G2\,V (\acen\,A) and K1\,V star (\acen\,B) and of the nature of the mysterious and unidentified object U, then situated about 5\asec\ north of A. The discovery of  U was reported in the 2016 paper. 

In the next section, Sect.\,2, we will discuss the new observations and the data reduction, with particular emphasis on the calibrations. The results for \acen\,A and B are provided in Sect.\,3, which are discussed in Sect.\,4. In Sect.\,5, we turn our attention toward the U source. Finally, in Sect.\,6, we briefly present our main conclusions.

\section{Observations and data reduction}

The observations of Cycle\,2 (C2, ID 2013.1.00170.S) have already been described and analysed by \citet{liseau2015} and \citet{liseau2016} and will not be repeated here, unless required by the context, (see, e.g., the upper half of Table\,\ref{Antennae}). 

The Cycle\,4 (C4, ID 2016.1.00441.S) observations using the number of  antennas shown in the lower half of Table\,\ref{Antennae} were performed during 2016 and 2017. Due to phase errors in the Band\,3 and 4 observations, these data had to be re-acquired, which resulted in the delayed access to the data. For all observations, the number of telescopes was larger during C4 and, therefore, a higher sensitivity, by about 25\% to 40\%, could be expected. The parameters of the synthesised beams are also provided in that table, where $a$ and $b$ refer to the major and minor axes, respectively, of the elliptical Gaussian beam at half power.

\begin{table}[h]
\caption{Instrumental parameters for Cycles 2 and 4, respectively}             
\label{Antennae}      
\centering          
\begin{tabular}{ ccccr}     
\hline\hline \\ 
\smallskip                                                                                                     
Cycle &  Obs. Date  &  $N$ 		& Synth. Beam 			& PA 		\\				
Band &  yyyy-mm-dd &  Ant.   		& $a \times b$ (\asec)$^2$& (\adeg )	\\
  \hline
  \noalign{\smallskip}		
C2	&			&			&			&					\\
3	& 2014-07-03	& 30			& $1.81 \times 1.22$		&  19		  	\\
7	& 2014-07-07	& 32			& $ 0.43\times 0.28$		&  47		  	\\
9	& 2014-07-18	& 31			& $0.22\times 0.16$		& 36		  	\\
6	& 2014-12-16	& 35			& $1.64\times 1.07$		& 71	  		\\
4	& 2015-01-18	& 34			& $3.16\times 1.67$		& 82		 	 \\
8	& 2015-05-02	& 37			& $0.77\times 0.68$ 		& $-70$		 \\
  \noalign{\smallskip}
  \hline
  \noalign{\smallskip}	
C4  	&			&			&						&		\\
4	& 2016-12-19	& 42			& $1.38 \times 1.01 $ 		& $-25$	\\
6	& 2016-12-30	& 44			&  $0.81 \times 0.70$ 		& $-50$	\\
9	& 2017-03-22	& 42		 	& $0.67 \times 0.51$ 		& 60		\\
7	& 2017-03-26	& 43			& $1.28 \times 1.08$			& 41		\\
8	& 2017-04-07	& 43 			& $1.08 \times 0.97$   		& 51		\\
3	& 2017-05-08	& 47			& $0.73 \times 0.50$  		& $-28$    	\\
\hline
\end{tabular}
\end{table}

\begin{table}
\caption{ALMA calibration for Cycles 2 and 4}             
\label{cal}      
\centering          
\begin{tabular}{ c c  clll}     
\hline\hline \\ 
\smallskip  
\smallskip                                                                                                       
    &  Obs. Date  &  \multicolumn{3}{c}{Calibration: Sources and Functionality} 	\\ 				
\cline{3-5}
\\     
 	  &  yyyy-mm-dd  &  Phase 	& Bandpass  	& Flux 			 \\	
  \hline
  \noalign{\smallskip}		
C2	&			&			&			&	\\
Bd	&			& 			&			 &				\\ 
3	& 2014-07-03	& J1617-5848	& J1427-4206	& Ceres		  	\\
7	& 2014-07-07	& J1617-5848	& J1427-4206	& Titan		  	\\
9	& 2014-07-18	& J1617-5848	& J1508-4953	& Ceres		  	\\
6	& 2014-12-16	& J1408-5712	& J1427-4206	& J1427-421	  	\\
4	& 2015-01-18	& J1617-5848	& J1617-5848	& Ceres		 	 \\
8	& 2015-05-02	& J1617-5848	& J1427-4206	& Titan		 	\\
  \noalign{\smallskip}
  \hline
  \noalign{\smallskip}	
C4  	&			&			&			&	\\
Bd	 & 			& 			&			&				\\ 
4	& 2016-12-19	& J1424-6807	& J1617-5848	& Ganymed	 	\\
6	& 2016-12-30	& J1424-6807	& J1617-5848   & Callisto			\\
9	& 2017-03-22	& J1424-6807 	& J1266-0547   & Titan			\\
7	& 2017-03-26	& J1427-4206	& J1424-6807	& Titan 			\\
8	& 2017-04-07	& J1424-6807	& J1617-5848   & Callisto			\\
3	& 2017-05-08	& J1424-6807	& J1617-5848  	& J1617-5848    	\\
\hline
\end{tabular}
\end{table}

The selection of the ALMA configuration, i.e. the maximum extent of the baselines, should ensure that the stars, having apparent diameters less than 10\,mas, would remain spatially unresolved at all observing frequencies. As a result, the observation of point sources should render the reduction of the interferometric data relatively straightforward, using a source model corresponding to the synthesised elliptical Gaussian telescope beam to obtain best-fits to the observed visibilities. 

\section{ALMA Data Calibration}

The visibilities were calibrated following standard procedures using the CASA package (Common Astronomy Software Application, version 5.1.1). The calibration sources and their functions during our 2014/15 and 2016/17 observing campaigns, respectively, are shown in Table\,\ref{cal}. For complex gain calibration and bandpass, quasars were generally used. These are listed in the table under ``Phase'' and ``Bandpass'', respectively. Flux calibration was achieved by mostly observing asteroids and moons in the solar system, but occasionally quasars were also used (see column ``Flux'' in Table\,\ref{cal}). 

The C2 observations were all re-reduced in the same way as those of C4. That should guarantee an overall homogeneous data set for both cycles (see, e.g., Table\,\ref{ceres}).

\begin{figure}[ht]
  \resizebox{\hsize}{!}{
  \rotatebox{00}{\includegraphics{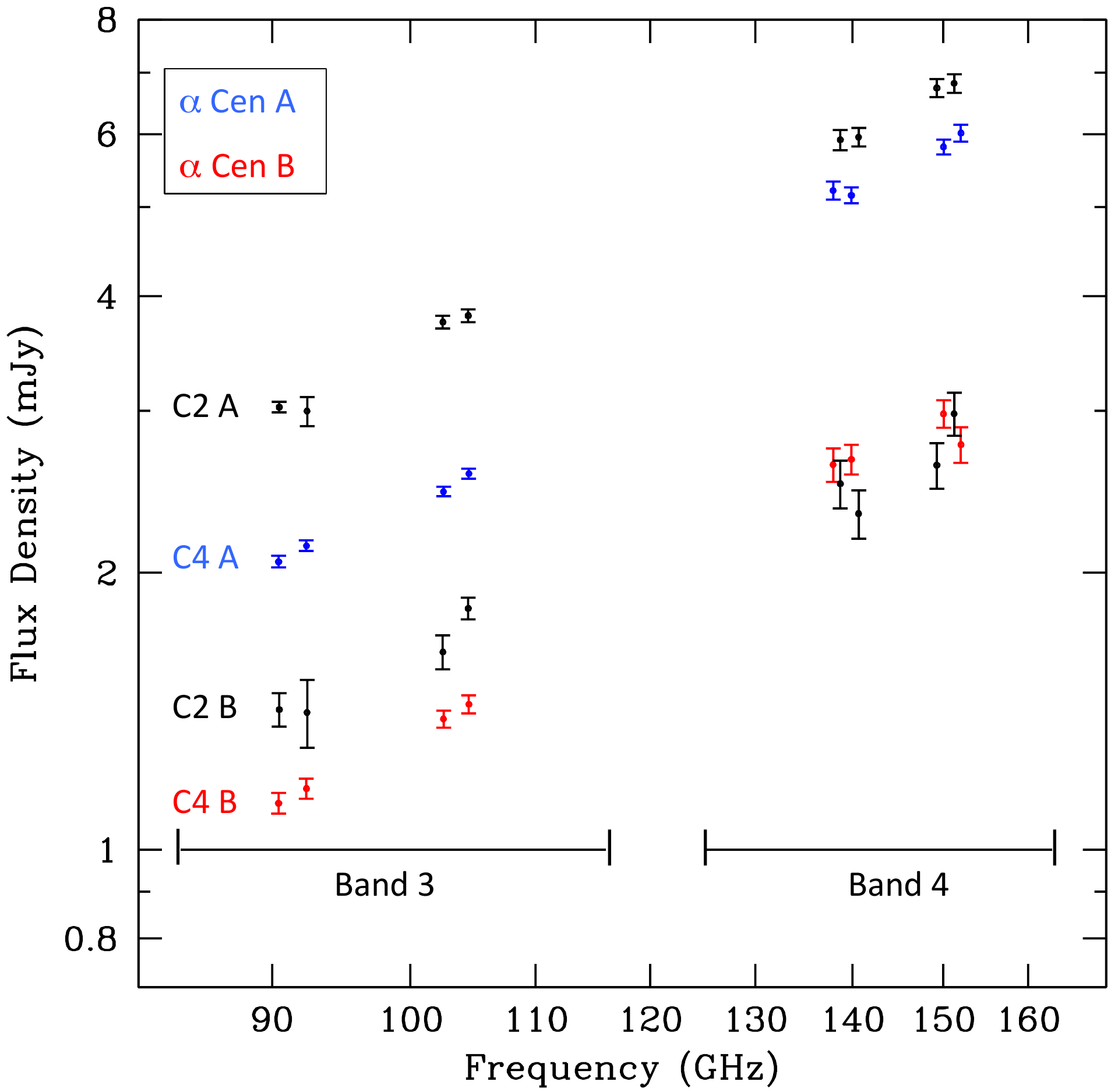}}
  }
  \caption{ALMA Band 3 and 4 flux densities of \acen A and B during two observing cycles (C2 and C4).  The four individual sub-band fluxes (2 GHz/spw) in a given band are shown, where black indicates C2 data of both stars. Re-calibrations are shown in blue for \acen A and in red for the cooler \acen B. Shown errors are statistical only. Regarding absolute levels, see the text (Sect.\,3.2).
  }
  \label{ALMA}
\end{figure}

\begin{table}
\caption{Cycle 2: Ceres calibration of  \acen A and \acen B}             
\label{ceres}      
\centering          
\begin{tabular}{ l l  }     
\hline\hline \\ 
\smallskip
\smallskip                                                                                                              
 Band 3:   $S_{97.5\,{\rm GHz}}$  (mJy)     	& Band 4:   $S_{145\,{\rm GHz}}$  (mJy)  \\         
\hline \\
Cycle 2                                                               &                                                 \\
A:  $3.37\pm 0.01$ [S/N=281]                    	&  $6.33\pm 0.08$ [S/N=83]   \\
B:  $1.59\pm 0.02$ [S/N=\phantom{1}80]     	&  $2.58\pm 0.08$ [S/N=34]  \\    
\hline    \\     
Cycle 4                                               	        &                                                 \\
A:  $3.159\pm 0.016$ [S/N=197]  		  	& $5.255\pm 0.037$ [S/N=142]     \\
 B: $1.549\pm 0.016$ [S/N=\phantom{1}97]    	& $2.481\pm 0.042$ [S/N=\phantom{1}59]  \\    
\hline
\end{tabular}
\end{table}

\subsection{Flux calibration with Ceres}

From the Cycle\,2 data, it came as a complete surprise that the spectral energy distribution (SED) appeared to change slope at the lowest ALMA frequencies, viz. in bands 3 and 4 \citep{liseau2016}. 

Based on the solar analogy, one would have expected this not to happen before at much lower frequencies, i.e. in the radio regime. In the paper by \citet[][Table\,2]{liseau2016}, the data for Bands 3 (2014-07-03) and 4 (2015-01-18) have been calibrated with an obsolete model for the asteroid Ceres. These observations took place before January 2015, and a Ceres model with a constant brightness temperature with frequency (185\,K) was used. Thereafter, thermo-physical models have been applied \citep[][see also: Appendix C in the CASA User Manual, https://casa.nrao.edu/docs/UserMan/casa\_cookbook014.html]{butler2012}. 

\subsection{Absolute flux calibration}

In order to determine absolute fluxes, the absolute systematic uncertainties need to be assessed and accounted for: currently, estimates and goals place these at better than 10\% in B3 (calibration from the quasar J1617-5848), and according to \citet{butler2012} better than 5-10\% in B4 (calibration from Ganymede), better than 10\% in B6 (calibration from Callisto), perhaps better than 20\% in B7 (calibration from Titan, no spectral lines),  better than 30\%  in B8 (calibration from Callisto) and maybe better than 30-50\%  in B9 (calibration from Titan, no spectral lines).

\begin{figure}
  \resizebox{\hsize}{!}{
    \rotatebox{00}{\includegraphics{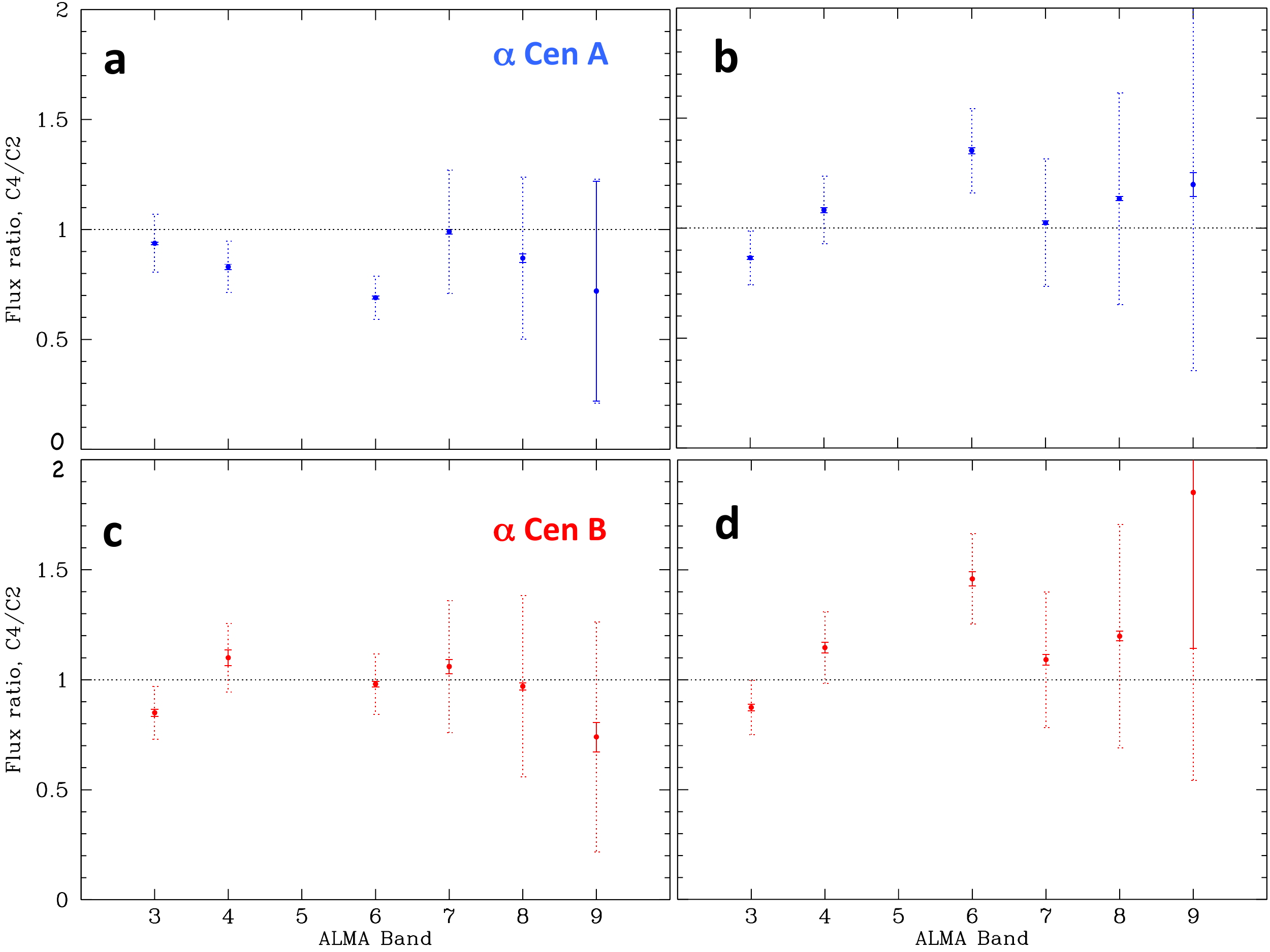}}
                        }
  \caption{Ratio of band-integrated flux densities obtained for \acen A (blue) during Cycle C2 in {\bf a} and  C4 in {\bf b}. Similar for \acen B (red) in {\bf c} and {\bf d}, respectively. In {\bf a} and {\bf c}, the C2 fluxes are those published by \citet{liseau2016}, whereas in {\bf b} and {\bf d}, the C2 fluxes are the re-reduced ones, i.e. those of Tables\,\ref{posA} and \ref{posB}. The dotted lines are based on the estimates of the absolute uncertainties according to \citet{butler2012}, and computed for independent measurements with individual errors $\partial S_{\!\nu}$, i.e. 
$\Delta R_{\nu}  =  \pm \,R_{\nu}  \sqrt{ (\partial S_{\!\nu,\,C4}/ S_{\!\nu,\,C4} )^2  +   (\partial S_{\!\nu,\,C2}/ S_{\!\nu,\,C2} )^2   }$, where  $R_{\!\nu} \equiv S_{\!\nu,\,C4}/S_{\!\!\nu,\,C2}\,$.
		}
  \label{C4_C2}
\end{figure}

The comparison of the absolute flux densities derived for the C4 data with those of C2 (both previous and current) is shown in Fig.\,\ref{C4_C2}.

\section{Results}

Also in Cycle\,4, both stars, \acena\ and \acenb, were detected at high signal to noise. As expected, the reconstructed stellar images were those of point sources outlining the synthesized elliptical telescope beams at the different frequencies. The adopted flux calibrations resulted in the values displayed in Tables\,\ref{posA} and \ref{posB}.

For \acen\,A, the results are presented in Table\,\ref{posA} which contains the data for both the re-reduced C2 and C4, ordered chronologically. The first column, designated Bd, shows the ALMA band identification, the frequency of which is given in column two.The Gregorian observing date and the Modified Julian Date follow, where MJD =  JD - 2400000.5. The measured J2000.0 equatorial coordinates (ICRS) with their error estimates are listed in columns 5 to 8. Finally, the flux densities and their statistical errors are found in columns 9 and 10. The corresponding information for \acen\,B is given in Table\,\ref{posB}.

Prior to 2015, Band\,9 observations (2014-07-18) were also flux calibrated using Ceres and also showed over-estimated flux densities compared to those taken in 2017-03-22. However, given the absolute uncertainties of up to 30-50\% in Band\,9 (calibration from Titan), this mismatch is formally within the errors. 

\begin{table*}
\caption{ALMA observations of \acen\,A during 2014 - 2017}             
\label{posA}      
\centering          
\begin{tabular}{ ccc ccc ccr c }     
\hline\hline \\ 
\smallskip  
\smallskip
Bd&    $\nu$	&   Obs.  Date	&   MJD      & RA (J2000)   &$\delta_{\rm RA}$ & Dec (J2000)                              & $\delta_{\rm Dec}$   & $S_{\nu}$ & $\delta\,{S_{\nu}}$ \\ 
    &	GHz	         &  yyyy  mm  dd&JD - 2400000.5&14\,h 39\,m + s &                s             & $-(60$\adeg\ 49\amin\ $+$\asec)  &    \asec                     & mJy	       & mJy    	   \\    
3  &	\phantom{1}91.487&2014 07 03& 56841.0557552&28.800&0.0004& 58.14    &    0.003  &   3.159   & 0.016 \\
7  & 	337.487	&  2014 07 07	& 56845.1131701 & 28.802  &  0.0003  &  58.18    &    0.002  &  25.989   & 0.191 \\
9  & 	675.002	&  2014 07 18	& 56856.0551458 & 28.777  &  0.0004  &  58.08    &    0.003  &  76.945   & 1.647 \\
6  & 	224.992	&  2014 12 16	& 57007.4699005 & 28.716  &  0.0008  &  57.61    &    0.004  &   9.373   & 0.082 \\
4  & 	138.987    &  2015 01 10	& 57040.5804693 & 28.663  &  0.0014  &  57.99    &    0.006  &   5.255   & 0.042 \\
8  & 	398.987    &  2015 05 02	& 57144.1387865 & 28.455  &  0.0003  &  58.29    &    0.002  &  30.850   & 0.175 \\
4  &  	138.987    &  2016 12 19	& 57741.4962025 & 27.654  &  0.0005  &  56.78   &    0.004  &   5.685   & 0.037 \\ 
6  &  	224.992    &  2016 12 29	& 57752.4800353 & 27.636  &  0.0003  &  56.79   &    0.002  &  12.675   & 0.063 \\
9  &  	675.002    &  2017 03 21	& 57834.3408501 & 27.511  &  0.0007  &  57.42   &    0.005  &  92.226   & 1.542 \\
8  &  	398.987    &  2017 03 26	& 57838.2734034 & 27.499  &  0.0004  &  57.46   &    0.003  &  34.969   & 0.188 \\
7  & 	337.487    &  2017 03 26	& 57838.2947216 & 27.498  &  0.0003  &  57.38   &    0.002  &  26.608   & 0.075 \\ 
3  & 	\phantom{1}91.487	&  2017 05 07	& 57881.1324039 & 27.374  &  0.0003  &  57.41   &    0.003  &   2.652   & 0.019 \\
 \hline
\end{tabular}
\end{table*}

\begin{table*}
\caption{ALMA observations of \acen\,B during 2014 - 2017}             
\label{posB}      
\centering          
\begin{tabular}{ ccc ccc ccr c }     
\hline\hline \\ 
\smallskip  
\smallskip
Bd&    $\nu$	&   Obs.  Date	&   MJD     & RA (J2000)   &$\delta_{\rm RA}$ & Dec (J2000)                              & $\delta_{\rm Dec}$   & $S_{\nu}$ & $\delta\,{S_{\nu}}$ \\ 
    &	GHz	        &  yyyy  mm  dd	&JD - 2400000.5&14\,h 39\,m + s &                s             & $-(60$\adeg\ 49\amin\ $+$\asec ) &    \asec                     & mJy	       & mJy    	   \\    
3  &	\phantom{1}91.487 &  2014 07 03	& 56841.0557552 & 28.252  &  0.0004  &  57.14   &    0.003   &   1.549   & 0.016 \\
7  & 	337.487	&  2014 07 07	& 56845.1131701 & 28.240  &  0.0003  &  57.28   &    0.002  &  11.322   & 0.248 \\
9  & 	675.002	&  2014 07 18	& 56856.0551458 & 28.216  &  0.0004  &  57.16  &    0.003  &  23.150   & 0.825 \\
6  & 	224.992	&  2014 12 16	& 57007.4699005 & 28.743  &  0.0008  &  56.41   &    0.004  &   4.171    & 0.083 \\
4  & 	138.987    &  2015 01 10	& 57040.5804693 & 28.163  &  0.0014  &  56.44   &    0.006 &    2.481   & 0.042 \\
8  & 	398.987    &  2015 05 02	& 57144.1387865 & 27.943  &  0.0003  &  56.80   &    0.002  &  13.404   & 0.188 \\
4  &  	138.987    &  2016 12 19	& 57741.4962025 & 27.259  &  0.0005  &  53.85   &    0.004 &   2.844     & 0.035 \\ 
6  &  	224.992	&  2016 12 29	& 57752.4800353 & 27.244  &  0.0003  &  53.83   &    0.002  &  6.089     & 0.056 \\
9  &  	675.002    &  2017 03 21	& 57834.3408501 & 27.134  &  0.0007  &  54.26   &    0.005  &  42.870   & 1.534 \\
8  &  	398.987    &  2017 03 26	& 57838.2734034 & 27.125  &  0.0004  &  54.30   &    0.003  &  16.050   & 0.188 \\
7  & 	337.487	&  2017 03 26	& 57838.2947216 & 27.124  &  0.0003  &  54.24   &    0.002  &  12.352   & 0.076 \\ 
3  & 	\phantom{1}91.487	&  2017 05 07	& 57881.1324039 & 27.009  &  0.0003  &  54.16   &    0.003    & 1.354   & 0.019 \\
 \hline
\end{tabular}
\end{table*}

\subsection{The SEDs of \acen A and B}

\begin{figure}[ht]
  \resizebox{\hsize}{!}{
  \rotatebox{00}{\includegraphics{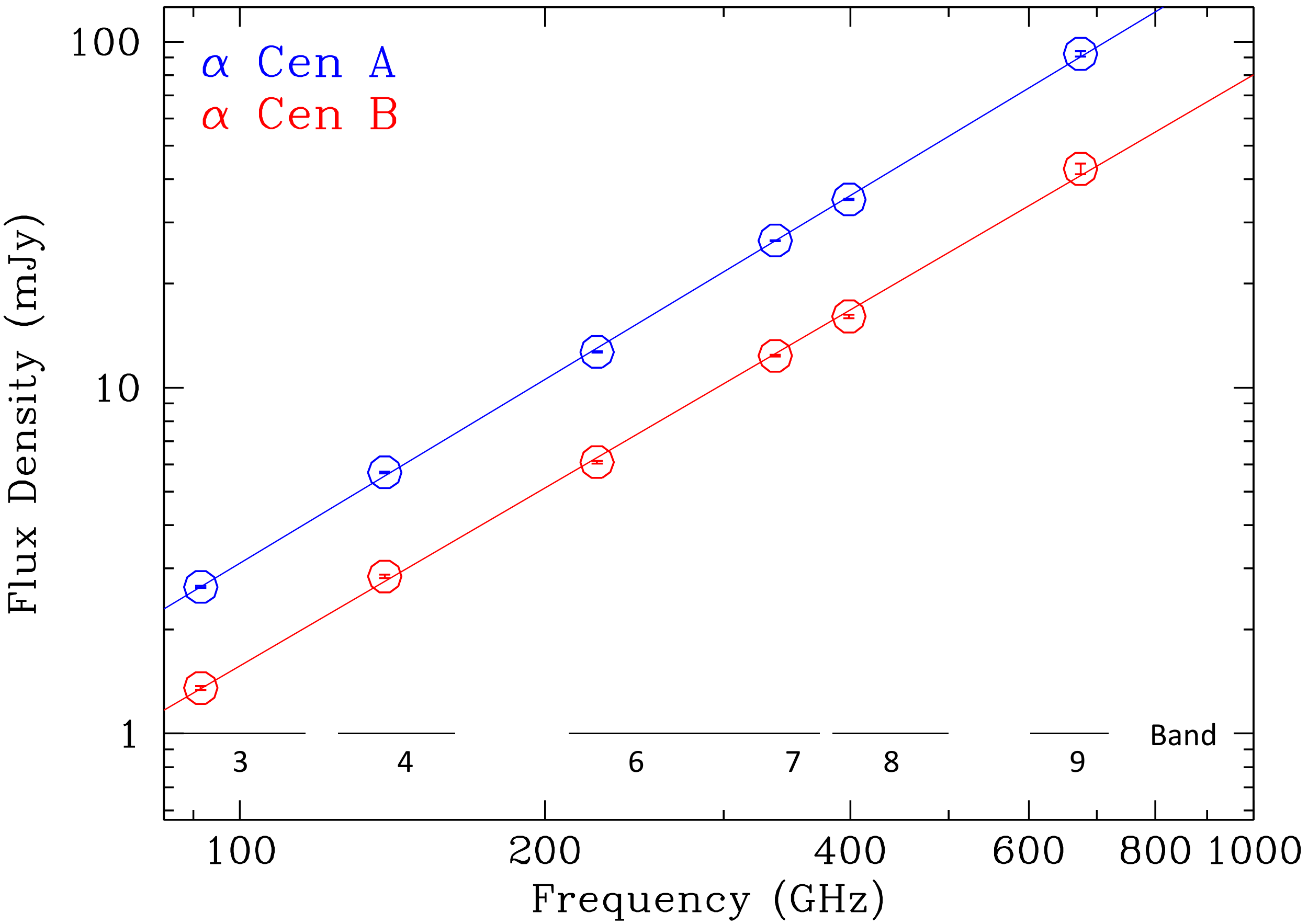}}
  }
  \caption{Band integrated flux densities in mJy as function of the frequency in GHz of \acena\ (blue symbols) and \acenb\ (red symbols) obtained during Cycle\,4. These were integrated over the ALMA bands, which are indicated below. The straight lines are fits to the logarithmic data and have slopes of about 1.7 (see Sect.\,4.1). The data points and their errors, which result from the fitting of the observed visibilities, are shown inside the large circles.
  }
  \label{SED}
\end{figure}

At the lower frequencies, the SEDs follow essentially the same trend as that determined at  higher frequencies, i.e. exhibiting $S_{\!\nu}$\,\about\,$\nu^{\,\alpha}$ over nearly two orders of magnitude in flux density (cf. Fig.\,\ref{SED}). For instance, $\alpha$ equals two for blackbody, i.e. optically thick, radiation in the Rayleigh-Jeans regime.

Linear regression \citep{press1986} of the C4 Band\,3 to 9 data results in $d \log S_{\!\nu}/d \log \nu = 1.76 \pm 0.01$ for \acen A and $1.71 \pm 0.02$ for \acen B, respectively (Fig.\,\ref{SED}). The SEDs appear flatter than those of blackbodies and also of what was obtained before. However, the results of the 2016 paper seem less significant. Clearly, with the current uncertain status of the absolute intensity calibration, our monitoring can at present not be used to meaningfully assess the level of stellar variability of \acen\ in the submm (see Fig.\,\ref{C4_C2}).

\subsection{The sky motions of the stars \acen A and B}

\begin{figure}
  \resizebox{\hsize}{!}{
  \rotatebox{00}{\includegraphics{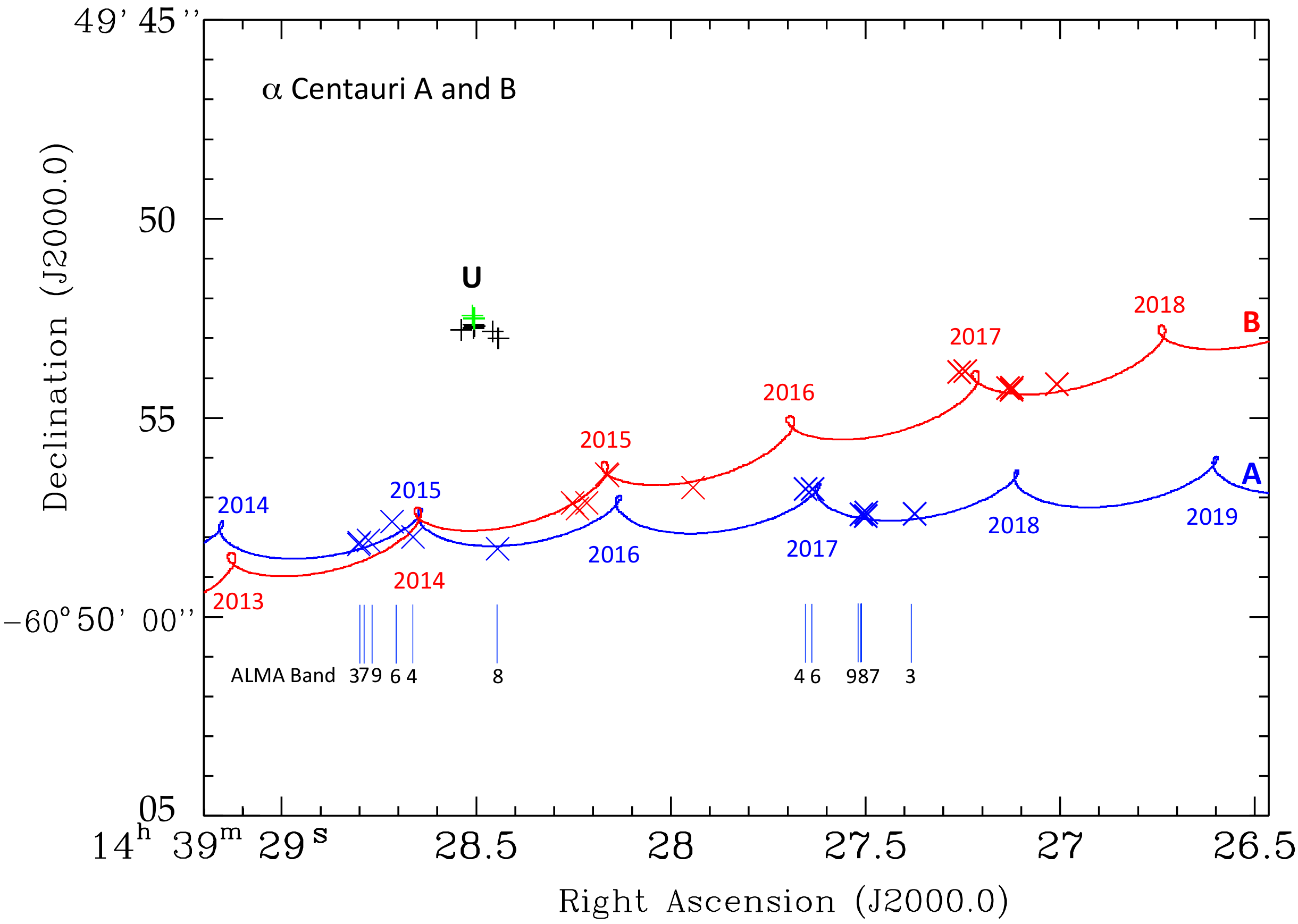}}
  }
  \caption{The displayed $20^{\prime \prime} \times 20^{\prime \prime}$ region shows the ephemerides of the \acen\ binary during the two ALMA-observing epochs. \acena\ is shown in blue and B in red. For clarity, the symbols for the observed positions in 2014 to 2017 are much larger than the statistical errors, which are on the milli-arcsecond scale. The ALMA bands for the positions of A are identified below the blue curve. Above both  curves, the measured sky positions of the U source are shown as crosses: in black for C\,2 (2014/15) and in green for C\,4 (2016/2017).}
  \label{ABU_ephem}
\end{figure}

The positional measurements of the components of the \acen\ binary, together with the graphical representation of their ephemerides, are shown in Fig.\,\ref{ABU_ephem}. These data are also provided in Tables \ref{posA} and \ref{posB}. The ephemerides were calculated adopting the stellar data from \citet{kervella2016}. Included in the ephemeris computations are the contributions from the orbital motions, the annual parallaxes and the proper motions of the binary stars, leading to complex patterns as shown in Fig.\,\ref{ABU_ephem}. The agreement of the theoretical results with the observational data is generally satisfactory.

\section{The nature of the U source}

\subsection{The ephemerides of U}

In Fig.\,\ref{ABU_ephem} also the corresponding data for the U source are shown, in black for the C\,2 and in green for the C\,4 observations. The source was firmly discovered in Band\,8 in 2015, only some 5\asec\ north of \acena\ \citep{liseau2016}. At the distance of the solar sibling, this would correspond to a projected orbit midway between Jupiter and Saturn in the solar system. As such, U seemed very intriguing and caught our interest.

Mainly because of detector saturation by the bright \acen, this  anonymous object had not been noticed before at any wavelength and its nature was thus undetermined, i.e. whether it was physically associated with \acen\ or an object in the fore- or background. Consequently, the object was termed U, meaning unidentified. To gain insight, data from at least another epoch would be required.
 
\begin{figure}
  \resizebox{\hsize}{!}{
  \rotatebox{00}{\includegraphics{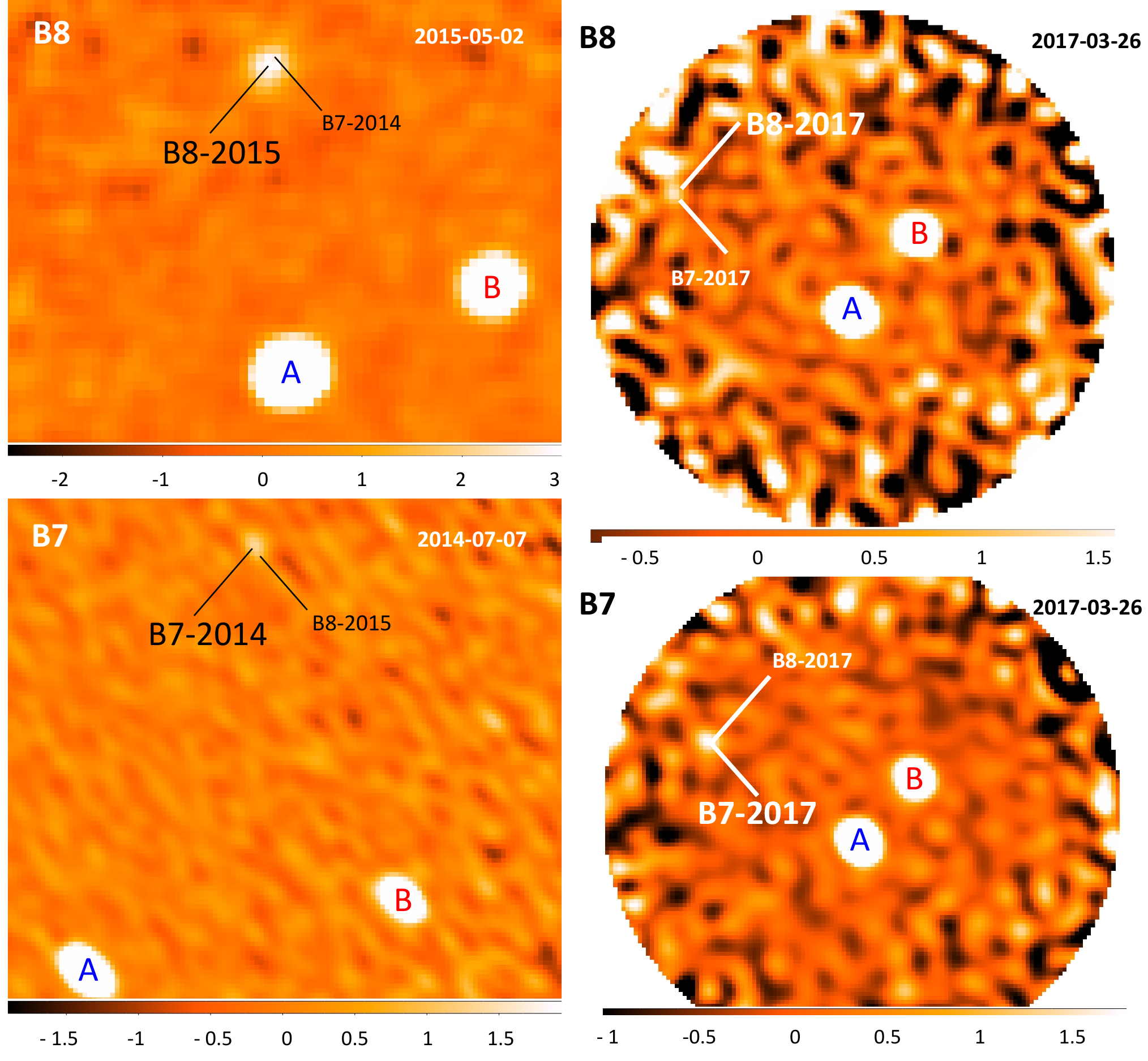}}
  }
  \caption{{\bf Left:} The relative positions of unidentified sources in the Bands 8 and 7 in the observed C\,2 images. B8-2015 is the discovery image of the U source, and the position of a corresponding object in the Band\,7 observation is indicated  (B7-2014). The spatial scale of B\,8 is given by the angular distance of \acen A and B, i.e.  $\overline{\rm AB}$ = \asecdot{4}{0}.  Below, part of the Band\,7 image, where the object B7-2014 is identified, together with the one of the upper frame (B8-2015). In that image,  $\overline {\rm AB}$ = \asecdot{4}{2} and the units of the intensity, shown along the color scale bar, are mJy/beam. North is up and East is to the left.
{\bf Right:} Similar to above, but for the C\,4 observations. The Band\,8 and 7 data have been obtained on the same date (2017-03-26) within half an hour. Consequently, the images of the U source coincide in both frames, as expected. The angular scale is given by $\overline {\rm AB}$ = \asecdot{4}{2}. The units of the intensity, shown along the color scale bar, are mJy/beam. North is up and East is to the left.  
  }
  \label{U_pos}
\end{figure}

In Table\,\ref{posU}, the data for the C\,2 and C\,4 observations are compiled. As previously mentioned, also during C\,4 was \acen\ observed in all bands. However, the U source would have been outside the primary beam of Band\,9 and, hence, no data are available for this source in that band. The identifications are based on positional coincidences across all the bands. In Figure\,\ref{U_pos}, the observations are shown for Bands\,8 and 7 in C\,2 (left) and C\,4 (right), respectively, demonstrating the reliability of this approach. On the right-hand side, the  Band\,8 and 7 images of 2017 are particularly compelling, as these data were taken on the same day and within 30 minutes: in both data sets is the U source situated exactly in the same place in the sky (see also Table\,\ref{posU}).

 \subsubsection{The proper motion of U}
 
The apparent sky motions of U with time do not allow any meaningful determination of its parallax. Its distance remains therefore unknown, but of course, it is much larger than that of \acen. The C\,2 data for Bands\,7, 9 and 8, and the C\,4 data for Bands\,6, 8 and 7 are collected around essentially the same Right Ascension and displaced by +\asecdot{0} {2} in Declination. Therefore, over the course of 1.9 years, any proper motion of U is limited to $\mu_{\alpha} < -$15\,mas\,yr$^{-1}$ and $\mu_{\delta} <$ 100\,mas\,yr$^{-1}$. These numbers essentially reflect the astrometric accuracy of the ALMA data. Anyway, these limits are very much different from the proper motion of \acen, i.e.  $\mu_{\alpha} = -$3620\,mas\,yr$^{-1}$ and $\mu_{\delta} =$ 694\,mas\,yr$^{-1}$ \citep{kervella2016}, and we conclude that U is (quasi-)stationary, implying that U is not part of the nearby high-proper-motion \acen\ system.

\begin{table*}
\caption{ALMA observations of the U source during 2014 - 2017l}             
\label{posU}      
\centering          
\begin{tabular}{ ccc cc  ll}     
\hline\hline \\ 
\smallskip  
\smallskip
Date		       & Bd	& $\nu$ 	    &  R.A.	 (2000)	&   Dec. (J2000)	    &	$S_{\nu}$   & $\delta S_{\nu}$  \\
yyyy mm dd     &	& GHz          &  hh mm ss.sss	&   \adeg\ \amin\ \asec\ & mJy		  & mJy		        \\			
2014\,\, 07 03  & 3 	&\phantom{1}97.5&14 39 28.458&  $-60$ 49 52.83 	    &	0.0554        & 0.0011 	 	\\
2015\,\, 01 18  & 4 	& 145	     &	14 39 28.539	&    $-60$ 49 52.79 	    &  0.1565        & 0.0015 		\\
2014\,\, 12 16  & 6 	& 233	     &	 14 39 28.444 	&    $-60$ 49 53.01 	    &  0.592   	  &  0.07  			\\
2014\,\, 07 07  & 7 	& 343.5	     &	14 39 28.507  	&    $-60$ 49 52.66 	    &  1.34		  & 0.4    			\\
2015\,\, 05 02  & 8 	& 405	     &	14 39 28.507  	&    $-60$ 49 52.74 	    &  3.2		  & 0.5   			\\ 
2014\,\,07 18   & 9 	& 679	     &	14 39 28.504 	&    $-60$ 49 52.70 	    &  6.7		  & 1.3  			\\ 
2016\,\, 12 30  & 6         & 233            & 14 39 28.510    &   $-60$ 49 52.42    &  0.559	  & 0.09     	 	\\
2017\,\, 03 26  & 8 	& 405	     &	14 39 28.505    &    $-60$ 49 52.50 	    &	1.47            & 0.6 			\\
2017\,\, 03 26  & 7 	& 343.5	     &	14 39 28.505    &    $-60$ 49 52.50 	    &	1.86            & 0.4			\\
2016\,\, 12 19  & 4 	& 145	     &	$\cdots$		&    $\cdots$		    &  $<0.08$      &$1\,\sigma$ 	\\		
2017\,\, 05 08  & 3  	&$\phantom{1}97.5$&$\cdots$	&   $\cdots$		    &	 $<0.017$	  &$1\,\sigma$		\\			         
 \hline
\end{tabular}
\end{table*}

\subsection{The SED of U}

\begin{figure}[ht]
  \resizebox{\hsize}{!}{
  \rotatebox{00}{\includegraphics{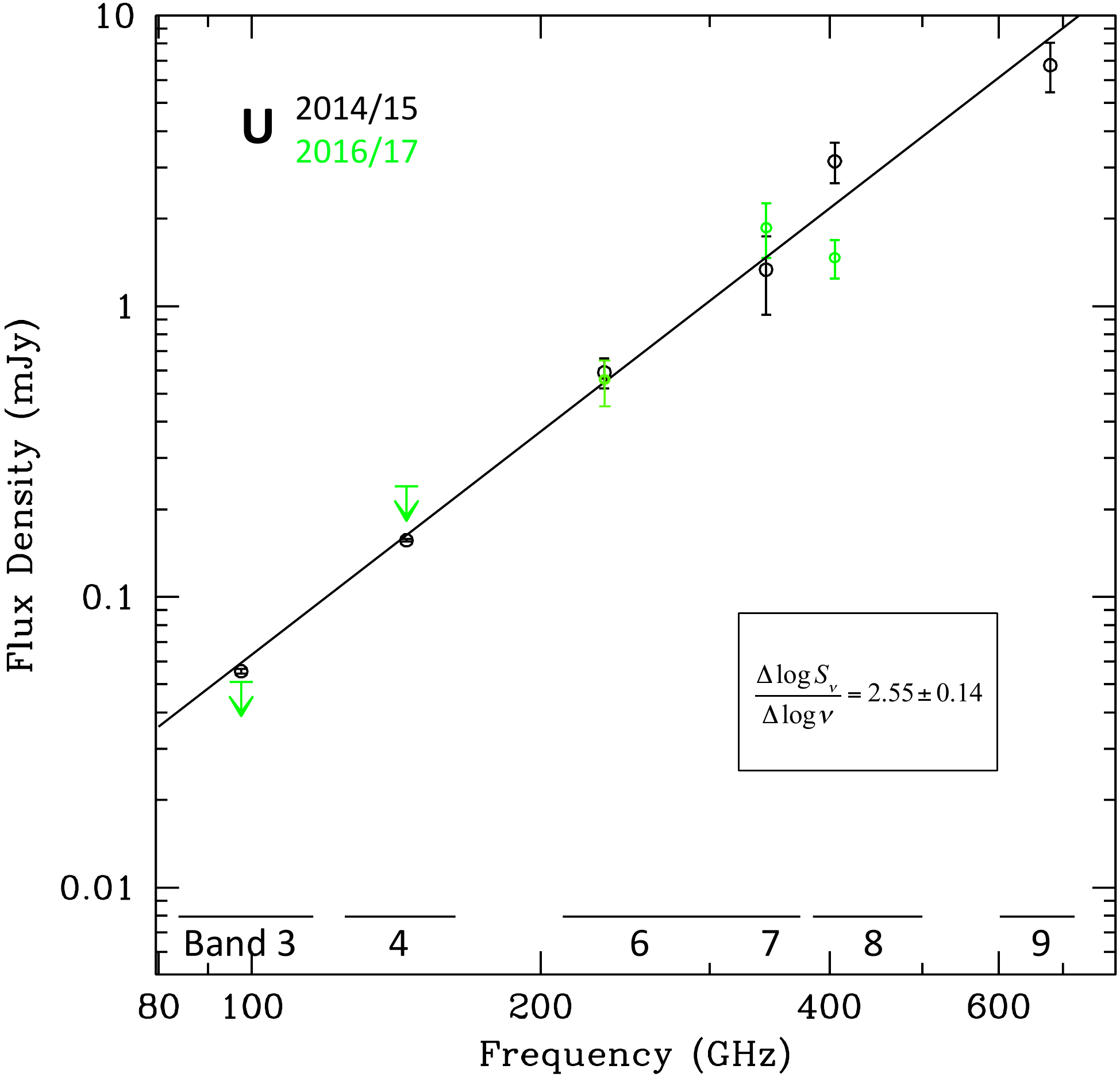}}
  }
  \caption{The observed SED with ALMA of the U source.  The C\,2 data are shown in black, whereas the C\,4 ($2016-2017$) data are in green. The upper limits are $3 \sigma$. The inset shows the results of a linear regression fit in log-log space.}
  \label{U_SED}
\end{figure}

As shown in Table\,\ref{posU}, flux data could be extracted for all bands during C\,2, whereas in C\,4, the source was not detected in Bands\,3 and 4 and, in Band\,9, it was outside the field of view. The resulting SED is shown in Fig.\,\ref{U_SED}, together with a regression fit to the logarithmic data. The slope of that line is $2.55 \pm 0.14$. If due to thermal emission by dust, an opacity exponent $\beta$ of about 0.5 would be implied, where $\kappa_{\nu} \propto \nu^{\,\,\beta}$. This is similar to the dust found in, e.g., the debris disc around \bpic\  \citep{liseau2003}, where the dust particles are significantly larger than those found in the diffuse interstellar medium, for which $\beta \sim 2$.

\section{Conclusions}

We have obtained 2nd epoch data for \acen tauri in ALMA-Bands 3, 4, 6, 7, 8 and 9 (92 to 675\,GHz). The Band\,3 data of ALMA Cycle\,4 had to be re-taken due to problems with the phase calibration, and that delayed the analysis. But the main reason for the delay was our discovery that the already published data (Cycle 2) for bands 3 and 4 were faulty: these were based on a flux calibration using Ceres that was eventually abandoned. Unfortunately, we were not made aware of that by the ALMA project, but found out in the course of the analysis of the Cycle 4 data. 

Here, we provide a complete re-reduction and analysis. With the aim of assessing the level of chromospheric time variability, the comparison of the results from the earlier C2 campaign with those from C4 casts doubt on the quality of the ALMA data even for the nearest stars to the Sun.

The mysterious object U that was discovered in May 2015, then about 5\asec\ north of \acena, has been re-observed during Cycle\,4 in all bands, except in Band\,9, where it fell outside the primary beam. These second epoch data were examined in order to establish whether U shared the proper motion of \acen. It does not. Hence, U is not related to the \acen\ system. Nevertheless, U appears to be an interesting object in its own right.


\begin{acknowledgements} We are grateful to Drs. Paresh Prema and Steve Bell at the HM Nautical Almanac Office, UK, for their kind help with essential data. We also wish to thank Dr. Sebastien Muller at the Nordic ALMA Regional Centre (ARC node) for his great help with the ALMA data. The Nordic ARC node is funded through Swedish
Research Council grant No 2017-00648. Dr. Dirk Petry at the ALMA User Support at ESO assisted with the Ceres calibration issues. This paper makes use of the following ALMA data: ADS/JAO.ALMA\#2013.1.00170.S and \#2016.1.00441.S. ALMA is a partnership of ESO (representing its member states), NSF (USA), and NINS (Japan), together with NRC (Canada) and NSC and ASIAA (Taiwan), in cooperation with the Republic of Chile. The Joint ALMA Observatory is operated by ESO, AUI/NRAO, and NAOJ.
\end{acknowledgements}

\bibliographystyle{aa}
\bibliography{ALMA}

\end{document}